\documentstyle[twoside,fleqn,npb,epsfig]{article}
\newcommand{\AmS}{{\protect\the\textfont2
  A\kern-.1667em\lower.5ex\hbox{M}\kern-.125emS}}

\title{ Precision $W$-pair physics with the YFSWW3 and KoralW Monte Carlos
}

\author{M. Skrzypek
  \address{Institute of Nuclear Physics,
  ul. Radzikowskiego 152, 31-342 Cracow, Poland \\
  and CERN, TH Division, CH-1211 Geneva 23, Switzerland}
\thanks{Talk given by M. Skrzypek at 6th International Symposium on
Radiative Corrections, 8--13 September 2002, Kloster Banz, Germany.}
, 
S. Jadach
  \address{Institute of Nuclear Physics,
  ul. Radzikowskiego 152, 31-342 Cracow, Poland}, 
W. P\l{}aczek
  \address{Institute of Computer Science, Jagellonian University,
   ul. Nawojki 11, 30-072 Cracow, Poland}, 
 B.F.L. Ward 
  \address{Department of Physics and Astronomy,
  The University of Tennessee, Knoxville, TN 37996-1200, USA} 
and  Z. W\c{a}s 
  \address{Institute of Nuclear Physics,
  ul. Radzikowskiego 152, 31-342 Cracow, Poland\\
  and CERN, TH Division, CH-1211 Geneva 23, Switzerland}
   }

\begin{document}

\begin{abstract}
We present the recent developments in the precision studies of $W$-pair
and single-$W$ processes in $e^+e^-$ collisions achieved with the help
of the KoralW and YFSWW3 Monte Carlo generators. We focus on
the theoretical precision of the measurements of $M_W$ and anomalous
couplings on the example of $\lambda$ coupling. We present the
mechanism of running these two independent codes in the form of one
Concurrent Monte Carlo code. We describe also the extensions of KoralW
necessary to emulate the kinematical region of single-$W$ process.  

\vspace{0.2cm}
\noindent
{\normalsize \bf CERN-TH/2002-261, UTHEP-02-0903, September 2002}

\end{abstract}


\maketitle

\section{Introduction}

In this note we give a short summary of our recent results and
developments
in the area of precision $W$-pair physics performed with the help of
the Monte Carlo (MC) event generators KoralW
\cite{koralw:1995a,koralw:1995b,koralw:1998,koralw:2001}
 and YFSWW3 
\cite{yfsww2:1996,yfsww3:1998,yfsww3:1998b,yfsww3:2000a,yfsww3:2001}. 
We will
concentrate mostly on the LEP2 physics; however, some
results for the energy range of future linear colliders will 
be also discussed.

The LEP2 data analysis is gradually approaching its final, highest
precision stage \cite{radcor02:chierici}. 
The forecast accuracy of 1\% for the W-pair total
cross section, 30 MeV for $W$ mass or 0.01 for anomalous coupling
$\lambda$ as an example of anomalous vector-boson triple-gauge
couplings (TGCs), will be an impressive achievement. Also in the
area of single-$W$ production the experimental precision for 
the total cross section is expected to be as high as 7\%. 

Such a good experimental
performance immediately puts demands on the theoretical
calculations -- they should be accurate to at least 
half of the experimental precision. 
All kinds of theoretical corrections have been extensively studied
throughout the last years and we will not attempt to cover this vast subject
here. Any interested reader will find a review in the
Reports \cite{LEP2YR:1996,LEP2YR:2000}.
We will only mention here that some of the most difficult corrections
turned out to be the ${\cal O}(\alpha)$ electroweak (EW) ones. These
corrections have been included in the YFSWW3 code in the so-called
``double-pole approximation'' based on the expansion in the
$\Gamma_W/M_W$ parameter and keeping only the highest (doubly) resonant
contributions, all done in a gauge-invariant way \cite{stuart:1997}. 
Apart from YFSWW3, only 
one other MC code, the RacoonWW \cite{Denner:2000bj,Denner:2000bm}, 
is able to calculate these corrections.

The available studies in the literature of theoretical precision have been 
restricted to the total cross section \cite{LEP2YR:2000}, 
lacking any discussion 
on the other
basic observables such as $M_W$ or
TGCs. The attempts to naively extrapolate the error of
$\sigma_{WW}$ to the other observables (in the form of ``error bands''
for example) have failed, as other observables {\em do not} depend
so much on the overall normalization, but rather on the shape
of some differential distributions. For that reason, we decided to study
both $M_W$ \cite{kandy:mw} and TGCs \cite{kandy:lambda} in detail. 
The TGC study has been done in
collaboration with R.~Bruneli\`ere,
A.~Denner,
 S.~Dittmaier, S.~J\'ez\'equel, M.~Roth and D.~Wackeroth. 

We have also done a number of upgrades and extensions of our MC
codes. Among them the most important ones are: (1) the ``merge without
merging'' of KoralW and YFSWW3 into a one MC code on an 
{\em event-per-event} basis,
with the help of the novel technique of Concurrent Monte Carlo
\cite{koralw:2001}, and (2) the
improvements in KoralW in the domain of single-$W$ processes
\cite{singlew}, 
necessary to meet the experimental precision in this area.

\section{KoralW\&YFSWW3 -- the Concurrent Monte Carlo for $W$ physics
\cite{koralw:2001} }

We have developed two Monte Carlo codes for the $W$-pair physics: KoralW and
YFSWW3. The first of them, KoralW, includes the complete, 
fully massive matrix element for all four-fermion processes, generated by 
the GRACE package \cite{GRACE2} along with two independent and
efficient multi-channel algorithms for the four-fermion phase-space
integration \cite{skrzypek:2000}. The YFSWW3 includes the complete 
${\cal
O}(\alpha)$ electroweak corrections to the $W$-pair production process 
in the double pole approximation, based on the on-shell calculations of
Refs.\ \cite{fleischer:1989,kolodziej:1991}. 
Apart from that, the two programs share a number of
features such as the initial state radiation (ISR) in 
Yennie--Frautschi--Suura exponentiation 
up to ${\cal
O}(\alpha^3)$ LL with multiple photons with non-zero transverse
momenta, Coulomb correction or anomalous TGCs. Also the semi-analytical
program KorWan for $W$-pair production and decay is included in both
packages. 

To meet the high precision requirements of the LEP experiments the
corrections of both codes need to be available simultaneously, on an {\em
event-per-event} basis. The most obvious way of achieving it, by merging
the two codes at the level of source codes, is highly impractical. The
already big codes (few hundred thousand lines) would become even larger;
it would not be possible to maintain them separately, or the merging
would have to be repeated after each new release. 
Therefore we proposed a different
approach: the Concurrent Monte Carlo (CMC) \cite{koralw:2001}. 
In CMC both codes remain
completely independent, but they run simultaneously and communicate in real
time through the UNIX/LINUX facility called ``named pipes'' or ``FIFO
files''.  From the
technical point of view, there is no
need to modify any of the codes (apart from the changes necessary to
accommodate the physics requirements) as the FIFO files behave just like
ordinary ASCII files. A sample of such a setup is shown
in Fig.~1,
\begin{figure}[htb]
\begin{center}
\setlength{\unitlength}{1mm}
\epsfig{file=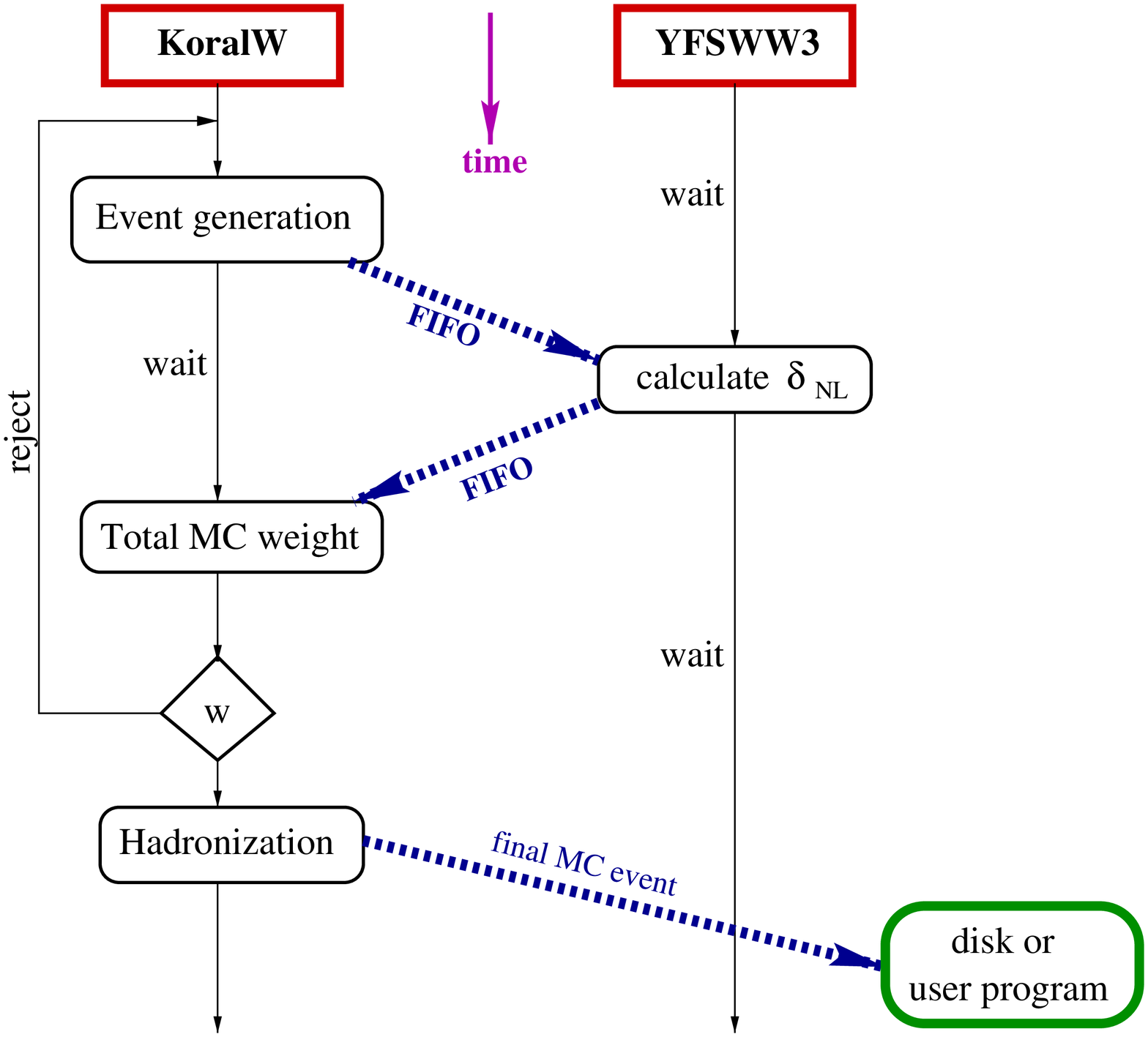,width=75mm,height=50mm}
\end{center}
\caption{The time-flow of the CMC KoralW\&YFSWW3}
\label{fig:kandy}
\end{figure}
where we present the main operation mode of CMC
KoralW\&YFSWW3
in which the events are constructed with full background
four-fermion corrections and the ${\cal O}(\alpha)$ EW corrections.

The concept of CMC has a number of advantages: (1) it provides
a natural mechanism of {\em ``encapsulation''} for any codes written in any
languages without any need to modify their source codes; (2) it has a
``lego block'' structure with many possible arrangements of involved 
codes
(for example for reprocessing data by two KoralW codes running
concurrently but initialized with {\em different} input parameters ); 
(3) it is naturally suited to use  
multiprocessor computers {\em efficiently}.

\section{Theoretical precision of \\ $M_W$ measurement \cite{kandy:mw}}

How did we proceed with the estimation of the theoretical error of the
$M_W$ measurement? We
set-up a simplified framework based on one-dimensional fits to the
invariant mass of one of the $W$'s. These fits were performed with
the help of the semi-analytical code KorWan \cite{koralw:1995b}, based on 
the CC03 matrix element convoluted with the ISR 
 and with the screened Coulomb correction, both matching
exactly the corresponding corrections in KoralW and YFSWW3 MCs.
In the first step, we checked that the fitting procedure works properly and
we reproduce in the fit the value of the $W$ mass used in the MC simulation. 
This introductory exercise shows also that if one is interested only in
the relative shifts of $M_W$ due to various effects, then any of the
fitting functions (FFs)
that we used is equally good.

In the next step we switched on and off various corrections in the MC sample
and in FFs to see how they influence the $M_W$. 
We found out that the most complicated corrections: 
four-fermion background 
and EW non-leading ${\cal O}(\alpha)$ ones 
are both of the size of 1 MeV. 

In the third step, we compared the results of the fits to the 
KoralW\&YFSWW3 data
with the fits to the data generated by the RacoonWW MC code, taken
from Ref.\ \cite{4f-LEP2YR:2000}. 
Both these codes include ${\cal O}(\alpha)$ EW
and four-fermion corrections, but they differ in many aspects of the
actual implementation of these corrections. 
Therefore such a comparison provides in particular a good test on 
the technical precision of the calculations.
The fitted value of $M_W$ obtained from corresponding data from these 
two codes turned out to differ by less than 3 MeV. 

Finally, we estimated the size of the possible missing corrections in
KoralW\&YFSWW3. This has been done for missing parts of ISR ($< 1$ MeV), FSR
($\sim 2$ MeV) as well as non-factorizable ($< 2$ MeV) ones. 

All the above work brought us to the following conclusion: the electroweak
theoretical precision of $M_W$ reconstruction through KoralW\&YFSWW3 at
LEP2 energies is $\sim$ 5 MeV. This estimate includes a ``safety factor'' 
of 2, due to the simplified 1-dimensional framework that we used. The above
estimate is in good agreement with the small differences between the YFSWW3 
and RacoonWW MCs. This value is well below the expected final LEP2
experimental precision of  $\sim$~30~MeV.

\section{Theoretical precision of anomalous $\lambda$ coupling
measurement \cite{kandy:lambda}} 

This study 
of $\lambda$ coupling has been done in collaboration with 
R.~Bruneli\`ere, A.~Denner,
S.~Dittmaier, S.~J\'ez\'equel, M.~Roth and D.~Wackeroth.  

The measurement of anomalous TGCs 
is primarily based on the distribution of the cosine of the $W$ boson 
production angle $\theta_W$. The most general parametrization of the 
anomalous TGCs requires fourteen complex parameters. In our analysis we
concentrated on the simplest scenario of only one  anomalous TGC, the
$\lambda= \lambda_\gamma=\lambda_Z$. We adopted essentially the same
strategy as for the $M_W$ analysis, i.e.\ we performed 1-dimensional
fits to the $\cos\theta_W$ distributions. 

As in the $M_W$ case, we began by a simple fit of the semi-analytical FF
from KorWan. It turned out, however, that the size of 
the various effects obtained 
from such a fit differ
significantly, depending on the acceptances applied to the MC
data. Therefore we had to find a way of incorporating more realistic
experimental acceptances into FFs. We did this by using the Monte Carlo
Parametric Fitting Function (MPFF),
constructed directly as a parametrization of MC data,  
in the form of a polynomial in
$\lambda$ and $\cos\theta_W$. In particular we used the real-life ALEPH
acceptances. The performed fits to MPFFs thus show that the 
${\cal O}(\alpha)$ NL EW corrections lead to a $\lambda$ shift of about
0.01 and that the differences between KoralW\&YFSWW3 and RacoonWW are
very small, below 0.001. In the next step we also looked at what
happens when we extend our analysis to the
different channels, other than $\mu\nu_\mu qq$. This has been done with
the help of ``transfer matrices'', constructed on the basis of the 
full 
ALEPH detector simulation. These matrices have been applied to our 
MC $\cos\theta_W$ distributions. As a result we showed that the effect
of ${\cal O}(\alpha)$ NL EW corrections on $\lambda$ is similar (within
15\%) in all channels, including the difficult four-quark and
tau-semileptonic ones. 
In the last step we have estimated the size of missing corrections in
both MC codes. 

As a result we came to the conclusion that the EW theoretical
uncertainty in $\lambda$ of the KoralW\&YFSWW3 and RacoonWW MC codes is 
$\sim 0.005$. As before this number includes a ``safety factor'' of 2
due to the 1-dimensional fit and missing higher-order effects.

\section{Improvements of KoralW for the single-$W$ processes
\cite{singlew} }

Apart from the $W$-pair production, another interesting process
involving the $W$ boson is the single-$W$ process. It is usually defined
as the $e\bar e \to e\bar\nu_e f'\bar f$ with outgoing $e$ 
close to the
beam line (say, $\cos\theta_e\geq 0.95$) \cite{4f-LEP2YR:2000}.
In such a configuration the $t$-channel graphs dominate. As a
consequence, the photonic radiation based on initial state only,
sufficient for $W$-pair production, 
has to be corrected to include the screening
of the electric charge (ECS) due to the dipole-like structure formed by
initial- and final-state electrons. 
This screening is responsible for the strong suppression of photons with
large transverse momenta and for the change of scale of 
the ``leading logarithms'' from $s$ to $|t|$. 

The KoralW code, based on ISR, has
missed the ECS effect. Therefore we have proposed \cite{singlew} 
a simple way of
reintroducing it by just adding an additional weight, without any 
modification of the original code. The weight results from 
an {\em educated guess} rather than the 
rigorous derivation from the first 
principles. It is constructed as a ratio of one-photon differential
distributions of a {\em coherent} sum of IS and FS radiation over an 
analogous {\em incoherent} IS and FS sum. These distributions are 
exact in the soft photon limit
and have the simple form of $(p_a/kp_a -p_b/kp_b)^2$. 
In order to
maintain the independence of the program of the dummy infra-red cut-off
$\epsilon_{IR}$ we also introduced the corresponding virtual weight. As
in the real-emission case, we have constructed it as an ansatz that
fulfills the requirement of cancelling the spurious $\epsilon_{IR}$
dependence. 
The overall precision of this ECS
ansatz we estimate at 2\%, because of some missing nonleading terms.

In order to meet the experimental requirements of LEP2, we had to
introduce also the effect of running the QED coupling constant in the
$e$ vertex from the $M_W^2$ to $t$ scale. We have done this in the  
form of an overall factor $(\alpha(t)/\alpha_{G_\mu})^2$. Such a naive 
ansatz has a
precision better than 2\%, see Ref.\ \cite{Passarino:2000mt}. 

In summary, we estimate at 3\% the overall precision of the proposed  
improvements in KoralW 
in the single-$W$ region of phase space. 
This is sufficient for the expected LEP2 experimental precision 
of about 7\%.

\begin{table*}[hbt]
\caption{The total cross sections with ECS effect (ECS) and running QED
coupling (RCC)}
\begin{tabular*}{\textwidth}{@{}l@{\extracolsep{\fill}}rrrr}
\hline
E [GeV] & $\phantom{\Big(}\sigma^{ECS}_{RCC}$ [fb]   
        & $1-\sigma_{RCC}/\sigma^{ECS}_{RCC}$ 
        & $1-\sigma^{ECS}/\sigma^{ECS}_{RCC}$ 
        & $1-\sigma^{ }/\sigma^{ECS}_{RCC}$ 
\\
\hline
190 & $87.11(22)$ 
    & $ 5.65(15)$
    & $-5.16( 2)$
    & $ 0.81(14)$
\\
200 & $103.60(26)$ 
    & $  5.49(15)$ 
    & $ -5.13( 2)$
    & $  0.67(15)$
\\
500 & $807.5(2.7)$ 
    & $  4.92(21)$
    & $ -4.68( 2)$
    & $  0.51(21)$
\\
\hline
\end{tabular*}
\end{table*}

We present in Table 1
some numerical results for the single-$W$ total
cross sections in the $e^+e^-\to e^-\bar\nu_e u \bar d$ process with the
following cuts: $\cos\theta_e\geq 0.997$, $M_{q\bar q}\geq 45$~GeV.
As can be seen, the effects of ECS and running of QED
coupling cancel each other in the total cross section in a
wide energy range, from the LEP2 up to future linear colliders area. 
This cancellation is however purely accidental and would not be
true for differential distributions.

\section{Summary}

In this short note we presented recent results and developments 
in precision $W$-pair
physics related to the KoralW and YFSWW3 MC programs: 

(1) 
    We
    established that the theoretical
    precisions, so far omitted in the literature, 
    of fundamental LEP2 observables other than total cross
    section are: 5 MeV for $M_W$ (LEP2 experimental precision $\sim$~30
    MeV) and
    0.005 for  anomalous $\lambda$ coupling (LEP2 experimental precision 
    $\sim$~0.01). These precisions fully meet the LEP2 experimental  
    requirements.

(2) We proposed a new way of combining independent MC codes on an {\em
    event-per-event} basis, by running them in parallel and 
    providing the real-time data exchange between them
    by means of the UNIX/LINUX facility of ``named pipes''. 
    We used this scheme to create a Concurrent MC
    KoralW\&YFSWW3 that generates $W$-pair events with both four-fermion
    background and ${\cal O}(\alpha)$ corrections. 

(3) We presented the upgrade of KoralW in the kinematical domain of
    single-$W$ targeted at better emulation of the 
    $t$-channel-dominated photonic radiation. 
    Thanks to this upgrade, the overall
    precision  of KoralW for this
    type of processes at LEP2 has become 3\% (LEP2 experimental
    precision $\sim$~7\%). 

\section*{Acknowledgements}

This work has been supported in part by the Polish Government 
grants KBN 5P03B09320, 2P03B00122, 
the European Commission 5-th framework contract HPRN-CT-2000-00149,
the NATO Grant PST.CLG.977751,
the US DoE contract DE-FG05-91ER40627
and the Polish-French Collaboration within IN2P3 through LAPP Annecy.


\end{document}